\documentclass[9pt,journal]{IEEEtran} 
\usepackage{amsmath}
\usepackage{amssymb}
\usepackage{multirow}
\usepackage{mathtools}
\usepackage{xcolor}
\usepackage{algorithm}
\usepackage{algorithmic}
\usepackage{tikz}
\usepackage{pgf-pie}
\usepackage{adjustbox}
\usepackage{caption}
\usepackage{mdframed}
\usepackage{hyperref}
\usepackage{bm}

\let\oldnorm\norm
\def\norm{\@ifstar{\oldnorm}{\oldnorm*}}
\makeatother

\usepackage{pifont}

\usepackage{array}

\ifCLASSOPTIONcompsoc
  \usepackage[nocompress]{cite}
\else
  \usepackage{cite}
\fi
\ifCLASSINFOpdf
\else
\fi

\begin{document}
%
\title{Cryptanalysis of \textsf{LC-MUME}: A Lightweight Certificateless Multi-User Matchmaking Encryption for Mobile Devices}

 \author{Ramprasad Sarkar
 	\IEEEcompsocitemizethanks{\IEEEcompsocthanksitem Ramprasad Sarkar is with the Cryptology and Security Research Unit, Indian Statistical Institute Kolkata, 203 B T Road, Kolkata, India-700108 \textit{(e-mail: rpsarkar123@gmail.com)}.\protect\\
 }
 }

%
%



\IEEEtitleabstractindextext{%
\begin{abstract}
Yang \textit{et al.} proposed a lightweight certificateless multi-user matchmaking encryption (\textsf{LC-MUME}) scheme for mobile devices, published in IEEE Transactions on Information Forensics and Security (TIFS) (DOI: 10.1109/TIFS.2023.3321961). Their construction aims to reduce computational and communication overhead within a one-to-many certificateless cryptographic framework. The authors claim that their scheme satisfies existential unforgeability under chosen-message attacks (\textsf{EUF-CMA}) in the random oracle model. However, our cryptanalytic study demonstrates that the scheme fails to meet this critical security requirement. In particular, we show that a \textsf{Type-I} adversary can successfully forge a valid ciphertext without possessing the complete private key of the sender. Both theoretical analysis and practical implementation confirm that this attack can be mounted with minimal computational cost. To address these weaknesses, we propose a modification strategy to strengthen the security of matchmaking encryption schemes in mobile computing environments.

\end{abstract}

\begin{IEEEkeywords}
Cryptanalysis, Identity-based encryption, Match-Making encryption, Chosen-ciphertext attack security, Anonymity
\end{IEEEkeywords}}

\maketitle

\IEEEdisplaynontitleabstractindextext
\IEEEpeerreviewmaketitle

\vspace{-0.5cm}

\section{Introduction}
\label{intro}
Matchmaking Encryption (\textsf{ME}) is an advanced cryptographic primitive that enables bilateral access control between the sender and the receiver. Unlike traditional encryption schemes that enforce access policies unilaterally, \textsf{ME} empowers the sender to define an access policy specifying which receivers may decrypt the message, while simultaneously allowing the receiver to verify whether the ciphertext originates from a legitimate sender. This dual control mechanism improves communication privacy, particularly by protecting the sender’s identity.

To address the shortcomings of conventional attribute-based encryption and signature-based mechanisms, Chen \textit{et al.}\cite{chen2021cl} introduced a certificateless matchmaking encryption (\textsf{CL-ME}) scheme tailored for IoT environments, presenting two efficient constructions based on bilinear pairings and lightweight cryptographic techniques. Building on this line of work, Yang \textit{et al.}\cite{yang2023lightweight} proposed a lightweight certificateless multi-user matchmaking encryption (\textsf{LC-MUME}) scheme for mobile platforms. Their construction aims to improve efficiency by avoiding pairings and relying on standard hardness assumptions. They claim that their scheme achieves existential unforgeability under chosen-message attacks (\textsf{EUF-CMA}).

\section{Review of Yang \textit{et al.}'s \textsf{LC-MUME}}
\label{review}
The original \textsf{LC-MUME} scheme includes five algorithms due to space limitations; here, we omit the detailed description, which can be referred to \cite{yang2023lightweight}.
 \vspace{-0.6cm}

\section{Our Proposed Attacks}
\label{forge-attack}
In this section, we identify the security vulnerabilities present in Yang et al.'s scheme \cite{yang2023lightweight}. A secure Certificateless Multi-User Matchmaking Encryption scheme should ensure that a sender cannot repudiate sending a valid encrypted message to a receiver. Additionally, it must prevent any adversary from impersonating the sender to create valid encrypted messages without knowing the full private key of the sender. Yang et al. \cite{yang2023lightweight} claim that their scheme is existentially unforgeable under a chosen message attack. However, we will demonstrate that a \textsf{Type-I} Adversary $\mathcal{ADV}_I$ can successfully forge a valid encrypted message to the receiver by substituting the sender's public key. The attack comprises the following three stages: 

\noindent\textbf{\underline{Step-1.}} In this stage, $\mathcal{ADV}_I$ replaces public key of the sender. For that, $\mathcal{ADV}_I$ randomly selects $a^{*},b^{*} \in \mathbb{Z}^{*}_q$ and replaces the corresponding public key $\textsf{PK}^{*1}_{\textsf{Id}_{\textsf{S}}} = a^{*}\textsf{P},$ $\textsf{PK}^{*2}_{\textsf{Id}_{\textsf{S}}} = b^{*}\textsf{P}$, while $\mathcal{ADV}_I$ uses the secret key for the sender as  $\textsf{SK}^{*}_{\textsf{Id}_{\textsf{S}}}=$ $(\textsf{SK}^{*1}_{\textsf{Id}_{\textsf{S}}}, \textsf{SK}^{*2}_{\textsf{Id}_{\textsf{S}}})=$ $(a^{*},b^{*})$. 

\noindent\textbf{\underline{Step-2.}} In this stage, $\mathcal{ADV}_I$ generates a ciphertext under the replaced public key $\textsf{PK}^{*}_{\textsf{Id}_{\textsf{S}}}$. $\mathcal{ADV}_I$ does the following.
\begin{itemize}
    \item[(i)] $\mathcal{ADV}_I$ chooses $r,s,d^{*}_1,d^{*}_2\in \mathbb{Z}^{*}_q$ and computes $\textsf{CT}^{*}_1=r\textsf{P}$ and $\textsf{CT}^{*}_2=s\textsf{P}-\mathcal{H}_1(\textsf{Id}_{\textsf{S}},\textsf{PK}^{*2}_{\textsf{Id}_{\textsf{S}}})\mathcal{P}^{'}$.
    
    \item[(ii)] For $\textsf{Id}_{\textsf{i}} \in \textsf{Rcvr}$, it computes 
    \begin{align*}
    \tau^{*}_i&=\textsf{SK}^{*1}_{\textsf{Id}_{\textsf{S}}}\cdot \textsf{PK}^{1}_{\textsf{Id}_{\textsf{i}}}\\
    \mathcal{V}^{*}_{\textsf{Id}_{\textsf{i}}}&=\mathcal{H}\left(r\cdot \mathcal{H}_4(\textsf{Id}_{\textsf{S}},\textsf{Id}_{\textsf{i}},\tau^{*}_i)\left(\textsf{PK}^{2}_{\textsf{Id}_{\textsf{i}}}+\mathcal{H}_1(\textsf{Id}_{\textsf{i}},\textsf{PK}^{2}_{\textsf{Id}_{\textsf{i}}})\mathcal{P}^{'}\right)\right)\\
    \mathcal{Z}^{*}_{\textsf{Id}_{\textsf{i}}}&=\left(s+ \textsf{SK}^{*2}_{\textsf{Id}_{\textsf{S}}}+\textsf{SK}^{*1}_{\textsf{Id}_{\textsf{S}}}\cdot\mathcal{H}_4(\textsf{Id}_{\textsf{S}},\textsf{Id}_{\textsf{i}},\tau^{*}_i)\right)\textsf{PK}^{1}_{\textsf{Id}_{\textsf{i}}}.   
    \end{align*} 
    
    \item[(iii)] Then the adversary sets two $n$-degree polynomials as follows.
    \begin{align*}
        f^{*}(x)&=\prod\limits_{\textsf{Id}_{\textsf{k}}\in \textsf{Rcvr}} (x-\mathcal{V}^{*}_{\textsf{Id}_{\textsf{k}}})+d^{*}_1= \sum\limits_{\textsf{k}=0}^{n-1} a^{*}_{\textsf{k}} x^{\textsf{k}} +x^{n} (\bmod ~q)\\
        g^{*}(y)&=\prod\limits_{\textsf{Id}_{\textsf{k}}\in \textsf{Rcvr}} (y-\mathcal{Z}^{*}_{\textsf{Id}_{\textsf{k}}})+d^{*}_2= \sum\limits_{\textsf{k}=0}^{n-1} b^{*}_{\textsf{k}} y^{\textsf{k}} +y^{n} (\bmod~q)       
    \end{align*}

    \item[(iv)] Then, it computes the following ciphertext components as follows.    
    \begin{align*}
        \textsf{CT}^{*}_3&=\left[\mathcal{H}_2(\textsf{CT}^{*}_1,\textsf{CT}^{*}_2,d^{*}_1,d^{*}_2)  \right]_{l-l_1} \parallel \left(\left[\mathcal{H}_2(\textsf{CT}^{*}_1,\textsf{CT}^{*}_2,\right.\right.
        \\
        &\left.\left.d^{*}_1,d^{*}_2)  \right]^{l_1}\bigoplus m\right) \\ \textsf{CT}^{*}_4&=\mathcal{H}_3(\textsf{CT}^{*}_1,\textsf{CT}^{*}_2,\textsf{CT}^{*}_3,a^{*}_0,a^{*}_1,\ldots,a^{*}_{n-1},b^{*}_0,b^{*}_1,\ldots,b^{*}_{n-1}).        
    \end{align*}

    \item[(v)] Finally, it returns a corresponding ciphertext $\textsf{CT}^{*}$= $\left(\textsf{CT}^{*}_1,\textsf{CT}^{*}_2,\textsf{CT}^{*}_3,\textsf{CT}^{*}_4, a^{*}_0,a^{*}_1,\ldots,a^{*}_{n-1},b^{*}_0,b^{*}_1,\right.$ $\left.\ldots,b^{*}_{n-1} \right)$.  
    
\end{itemize}
\noindent\textbf{\underline{Step-3.}} We notice that, since there’s no binding between a user’s identity and his public key, the receiver cannot detect that the sender's public key is replaced by the adversary. In this stage, upon receiving the ciphertext, the receiver $\textsf{Id}_{\textsf{i}}$ invokes the decryption algorithm as follows.
\begin{enumerate}
    \item It parses the ciphertext $\textsf{CT}^{*}= \left(\textsf{CT}^{*}_1,\textsf{CT}^{*}_2,\textsf{CT}^{*}_3,\right.$ $\left.\textsf{CT}^{*}_4, a^{*}_0,a^{*}_1,\ldots,a^{*}_{n-1},b^{*}_0,b^{*}_1,\ldots,b^{*}_{n-1} \right)$ and check whether the equation $\textsf{CT}^{*}_4=\mathcal{H}_3\left(\textsf{CT}^{*}_1,\textsf{CT}^{*}_2,\textsf{CT}^{*}_3,\right.$ $\left.a^{*}_0,a^{*}_1,\ldots,a^{*}_{n-1},b^{*}_0,b^{*}_1,\ldots,b^{*}_{n-1}\right)$.

    \item If not, returns $\bot$. Otherwise, it calculates \begin{align*}
        &\tau^{*}_{\textsf{i}}=\textsf{SK}^{1}_{\textsf{Id}_\textsf{i}} \cdot \textsf{PK}^{*1}_{\textsf{Id}_{\textsf{S}}},~
        \mathcal{V}^{*}_{\textsf{Id}_{\textsf{i}}}=\mathcal{H}(\textsf{SK}^{2}_{\textsf{Id}_\textsf{i}} \cdot \mathcal{H}_4(\textsf{Id}_{\textsf{S}},\textsf{Id}_{\textsf{i}},\tau^{*}_{\textsf{i}}) \cdot \textsf{CT}^{*}_1)\\
        &\mathcal{Z}^{*}_{\textsf{Id}_{\textsf{i}}}=\textsf{SK}^{1}_{\textsf{Id}_\textsf{i}}\left(\textsf{CT}^{*}_2+ \textsf{PK}^{*2}_{\textsf{Id}_{\textsf{S}}}+\mathcal{H}_1(\textsf{Id}_{\textsf{S}}, \textsf{PK}^{*2}_{\textsf{Id}_{\textsf{S}}})\mathcal{P}^{'}+\mathcal{H}_4\left(\textsf{Id}_{\textsf{S}},\textsf{Id}_{\textsf{i}},\right.\right.\\
        &\left.\left.\tau^{*}_i\right)\textsf{PK}^{1}_{\textsf{Id}_\textsf{i}}\right). 
    \end{align*}

    \item It then recovers $d^{*}_1$ and $d^{*}_2$ by computing $d^{*}_1=f(\mathcal{V}^{*}_{\textsf{Id}_{\textsf{i}}}), d^{*}_2=g(\mathcal{Z}^{*}_{\textsf{Id}_{\textsf{i}}})$. Then it finally returns the message as follows $m=\left[\mathcal{H}_2(\textsf{CT}^{*}_1,\textsf{CT}^{*}_2,d^{*}_1,d^{*}_2)  \right]^{l_1} \bigoplus \left[\textsf{CT}^{*}_3\right]^{l_1}$ if $\left[\mathcal{H}_2(\textsf{CT}^{*}_1,\textsf{CT}^{*}_2,d_1,d_2)  \right]_{l-l_1} = \left[\textsf{CT}^{*}_3\right]_{l-l_1},$ otherwise it return $\bot$.
\end{enumerate}

The adversary generated a forged ciphertext that is well-defined and correct due to the following calculations. \vspace{-0.4cm}

\begin{align*}
   \tau^{*}_i&=\textsf{SK}^{*1}_{\textsf{Id}_{\textsf{S}}}\cdot \textsf{PK}^{1}_{\textsf{Id}_{\textsf{i}}}=a^{*}x_{\textsf{i}}\textsf{P}\\
   \mathcal{V}^{*}_{\textsf{Id}_{\textsf{i}}}&=\mathcal{H}(\textsf{SK}^{2}_{\textsf{Id}_\textsf{i}} \mathcal{H}_4(\textsf{Id}_{\textsf{S}},\textsf{Id}_{\textsf{i}},\tau^{*}_{\textsf{i}}) \textsf{CT}^{*}_1)
   \\
   &=\mathcal{H}(r  \mathcal{H}_4(\textsf{Id}_{\textsf{S}},\textsf{Id}_{\textsf{i}},\tau^{*}_{\textsf{i}}) d_{\textsf{i}}\textsf{P})\\
   &=\mathcal{H}(r \cdot \mathcal{H}_4(\textsf{Id}_{\textsf{S}},\textsf{Id}_{\textsf{i}},\tau^{*}_{\textsf{i}}) (\textsf{PK}^{2}_{\textsf{Id}_{\textsf{i}}} + \mathcal{H}_1(\textsf{Id}_{\textsf{i}},\textsf{PK}^{2}_{\textsf{Id}_{\textsf{i}}}) \mathcal{P}^{'}))\\
   \mathcal{Z}^{*}_{\textsf{Id}_{\textsf{i}}}&=\textsf{SK}^{1}_{\textsf{Id}_\textsf{i}}\left(\textsf{CT}^{*}_2+ \textsf{PK}^{*2}_{\textsf{Id}_{\textsf{S}}}+\mathcal{H}_1(\textsf{Id}_{\textsf{S}},\textsf{PK}^{*2}_{\textsf{Id}_{\textsf{S}}})\mathcal{P}^{'}\right.\\
    &\left.+\mathcal{H}_4(\textsf{Id}_{\textsf{S}},\textsf{Id}_{\textsf{i}},\tau^{*}_i)\textsf{PK}^{*1}_{\textsf{Id}_{\textsf{S}}}\right)\\
    &=x_{\textsf{i}}\left(s\textsf{P}+b^{*}\textsf{P} +\mathcal{H}_4(\textsf{Id}_{\textsf{S}},\textsf{Id}_{\textsf{i}},\tau^{*}_i)a^{*}\textsf{P}\right)\\
    &=x_{\textsf{i}}\textsf{P} \left(s+b^{*}+\mathcal{H}_4(\textsf{Id}_{\textsf{S}},\textsf{Id}_{\textsf{i}},\tau^{*}_i)a^{*}\right)\\
    &=\left(s+ \textsf{SK}^{*2}_{\textsf{Id}_{\textsf{S}}}+\textsf{SK}^{*1}_{\textsf{Id}_{\textsf{S}}} \cdot\mathcal{H}_4(\textsf{Id}_{\textsf{S}},\textsf{Id}_{\textsf{i}},\tau^{*}_i)\right)\textsf{PK}^{1}_{\textsf{Id}_{\textsf{i}}}
\end{align*}\vspace{-0.5cm}

\noindent Since the receiver $\textsf{Id}_{\textsf{i}}$ is from the set $\textsf{Rcvr}$, therefore, it can recover the $d_1$ and $d_2$ by computing $d^{*}_1=f(\mathcal{V}^{*}_{\textsf{Id}_{\textsf{i}}}), d^{*}_2=g(\mathcal{Z}^{*}_{\textsf{Id}_{\textsf{i}}})$. This follows that the forged ciphertext $\textsf{CT}^{*}= \left(\textsf{CT}^{*}_1,\textsf{CT}^{*}_2,\textsf{CT}^{*}_3,\right.$ $\left.\textsf{CT}^{*}_4, a^{*}_0,a^{*}_1,\ldots,a^{*}_{n-1},b^{*}_0,b^{*}_1,\ldots,b^{*}_{n-1} \right)$ is valid. Therefore, the scheme is subject to universal forgery with respect to a \textsf{Type-I} adversary $\mathcal{ADV}_I$ who replaces the sender’s public key.

\section{Performance Analysis and Implementation} \label{perform}

This section presents a performance evaluation and practical implementation of the proposed forgery attacks against the certificateless matchmaking encryption scheme of Yang \textit{et al.}~\cite{yang2023lightweight}, under the \textsf{EUF-CMA} security model. In this setting, a \textsf{Type-I} adversary $\mathcal{ADV}_I$ attacks in two main stages: Step~1 and Step~2, ultimately generating a valid forged ciphertext. In Step~3, the forged ciphertext can be verified and correctly decrypted by an authorized user or the challenger, thus violating the \textsf{EUF-CMA} security notion.

The total cost incurred by $\mathcal{ADV}_I$ is the sum of all operations performed during Steps~1 through 3. Table~\ref{tab:combined} summarizes the computational complexity per operation step, as well as the total execution time (in milliseconds) required to complete the forgery for various values of the target user set size $n$.

Our attack technique was implemented on a Dell laptop equipped with an \textsf{AMD A9-9400 Radeon} processor, 12\,\textsf{GB} RAM, running Ubuntu 22.04.4 \textsf{LTS} (64-bit) with GNOME 42.9. We used the Pairing-Based Cryptography (\textsf{PBC}) library, version 0.5.14~\cite{lynn2013pbc}, utilizing a Type \textsf{A} bilinear pairing over the supersingular curve defined by $y = x^3 + x$.

The attack strategy was executed for different sizes of the user groups: $n \in \{8, 16, 32, 64, 128, 256, 512, 1024\}$. The timing results (in milliseconds) are reported in Table~\ref{tab:combined}, along with the detailed breakdown of the cryptographic operations involved. The findings indicate that the proposed attack is highly efficient and scales linearly with the number of target user sets.

\begin{table}[h]
\caption{Computation Cost and Execution Time of \textsf{EUF-CMA} Attack}
\label{tab:combined}
\begin{center}
\scalebox{0.68}{
\begin{tabular}{|c|c|c|c|c|c|c|c|c|c|c|}
\hline
\textbf{Step} & \textsf{\#Z} & \textsf{\#SM} & \textsf{\#SS} & \textsf{\#Hash} & \multicolumn{6}{c|}{\textsf{Attack Time (ms) for Varying User Sizes ($n$)}} \\ \cline{6-11}
 & & & & & 8 & 32 & 128 & 256 & 512 & 1024 \\ \hline
Step-1 & 2 & 2 & -- & -- & \multirow{3}{*}{11.20} & \multirow{3}{*}{39.04} & \multirow{3}{*}{150.40} & \multirow{3}{*}{298.88} & \multirow{3}{*}{595.84} & \multirow{3}{*}{1189.76} \\ \cline{1-5}
Step-2 & $2n+4$ & $3n+1$ & $7n+3$ & $n+4$ & & & & & & \\ \cline{1-5}
Step-3 & -- & -- & 7 & 6 & & & & & &\\ \hline
\end{tabular}
}
\end{center}
\scriptsize
\textsf{\#Z}: element generation in $\mathbb{Z}_p^{*}$, \textsf{\#SM}: scalar multiplication in the source or target group, \textsf{\#SS}: group addition/multiplication, \textsf{\#Hash}: hash function evaluation. $n$: number of users.
\end{table}

\section{Countermeasures Against the Proposed Attacks} \label{miti}

As shown in Section~\ref{forge-attack}, Yang \textit{et al.}’s scheme~\cite{yang2023lightweight} is vulnerable to forgery attacks launched by a \textsf{Type-I} adversary. The root cause of this vulnerability lies in unaccounted algebraic dependencies within the underlying group structure, which the security proof fails to capture. In particular, these attacks do not imply that the adversary solves the underlying hard problems in a general or random instance.

Attacks on certificateless encryption systems can broadly be classified into two categories:

\begin{itemize}
    \item \textbf{Passive Attacks:} These involve eavesdropping or traffic analysis, where the adversary observes the system's communication without altering its operation, aiming to extract useful information.
    
    \item \textbf{Active Attacks:} These are more intrusive, wherein the adversary interferes with system operations, e.g., by injecting, modifying, or replacing cryptographic elements, to breach security properties such as integrity or authenticity.
\end{itemize}

Our forgery attacks presented in this work fall into the category of active attacks. Specifically, the adversary replaces a user's public key to craft a valid forgery. These attacks are not only feasible but also practical, and thus must be addressed correctly in real-world deployments.

\noindent\textbf{\underline{Eliminating the \textsf{IB} Setup.}} A viable countermeasure is to eliminate the identity-based (\textsf{IB}) setup from the protocol. In an \textsf{IB} setting, the adversary can query key-generation oracles for arbitrary identities. This flexibility allows manipulation of public/private key relationships, which is exploited in the forgery attack. 

In contrast, certificateless matchmaking encryption schemes designed without an \textsf{IB} setup restrict the adversary’s access: key-generation and hash queries must reference opaque indices mapped to hidden identities. As a result, the adversary cannot associate public keys with real identities, rendering the attack ineffective.

However, the main trade-off is the loss of operational simplicity that \textsf{IB} setups provide, such as reduced certificate management and simplified trust models. Thus, while removing the \textsf{IB} setup offers better resistance to active forgeries, it may require careful design adjustments to preserve usability and scalability.

\vspace{-0.4cm}

\section{Conclusion}
\label{conclu}
We have performed a comprehensive cryptanalysis of the Lightweight Certificateless Multi-User Matchmaking Encryption scheme proposed by Yang \textit{et al.}, revealing critical security weaknesses. Specifically, our analysis shows that the scheme does not achieve unforgeability in a multi-user environment. To substantiate our findings, we present explicit attack scenarios that exploit these vulnerabilities, accompanied by a detailed evaluation of the associated computational costs. Furthermore, we propose a potential design strategy aimed at constructing secure and efficient matchmaking encryption protocols suitable for mobile computing framework.
\vspace{-0.4cm}

\ifCLASSOPTIONcompsoc
  
\else
  \section*{Acknowledgment}
\fi
This study was funded by the Information Security Education and Awareness (ISEA) Project Phase-III initiatives of the Ministry of Electronics and Information Technology (MeitY) under Grant No. F.No. L-14017/1/2022-HRD.
\vspace{-0.4cm}

\ifCLASSOPTIONcaptionsoff
  \newpage
\fi


\bibliographystyle{IEEEtran}
\bibliography{cas-refs}
\end{document}